\documentclass[aps,prd,twocolumn,superscriptaddress]{revtex4}
\usepackage{epsfig,epsf}
\usepackage{amsmath}
\usepackage{amsthm}
\usepackage{amsfonts}
\usepackage{amssymb}
\usepackage{dsfont}
\usepackage{multirow}
\usepackage{appendix}
\usepackage{slashed}
\usepackage[active]{srcltx}
\usepackage{psfrag}

\begin{document}

\title{Modeling the resonance $T_{cs0}^{a}(2900)^{++}$ as a hadronic
molecule $D^{\ast +}K^{\ast +}$ }
\date{\today}
\author{S.~S.~Agaev}
\affiliation{Institute for Physical Problems, Baku State University, Az--1148 Baku,
Azerbaijan}
\author{K.~Azizi}
\affiliation{Department of Physics, University of Tehran, North Karegar Avenue, Tehran
14395-547, Iran}
\affiliation{Department of Physics, Do\v{g}u\c{s} University, Dudullu-\"{U}mraniye, 34775
Istanbul, T\"{u}rkiye}
\author{H.~Sundu}
\affiliation{Department of Physics, Kocaeli University, 41380 Izmit, T\"{u}rkiye}
\affiliation{Department of Physics Engineering, Istanbul Medeniyet University, 34700
Istanbul, T\"{u}rkiye}

\begin{abstract}
The doubly charged scalar resonance $T_{cs0}^{a}(2900)^{++}$ is studied in
the context of the hadronic molecule model. We consider $%
T_{cs0}^{a}(2900)^{++}$ as a molecule $M=D^{\ast +}K^{\ast +}$ composed of
vector mesons, and calculate its mass, current coupling and full width. The
spectroscopic parameters of $M$, i.e., its mass and current coupling, are
found by means of the QCD two-point sum rule method by taking into account
vacuum expectation values of quark, gluon and mixed operators up to
dimension $10$. The width of the molecule $M$ is evaluated through the
calculations of the partial widths of the decay channels $M \to
D_{s}^{+}\pi^{+}$, $M \to D_{s}^{\ast +}\rho^{+}$, and $M \to D^{\ast
+}K^{\ast +}$. Partial widths of these processes are determined by strong
couplings $g_1$, $g_2$, and $g_3$ of particles at vertices $%
MD_{s}^{+}\pi^{+} $, $MD_{s}^{\ast +}\rho^{+}$, and $M D^{\ast +}K^{\ast +}$%
, respectively. We calculate the couplings $g_i$ by employing the QCD
light-cone sum rule approach and technical tools of the soft-meson
approximation. Predictions obtained for the mass $m=(2924 \pm 107)~\mathrm{%
MeV}$ and width $\Gamma=(123 \pm 25)~\mathrm{MeV}$ of the hadronic molecule $%
M$ allow us to consider it as a possible candidate of  the resonance $%
T_{cs0}^{a}(2900)^{++}$.
\end{abstract}

\maketitle

%%%%%%%%%%%%%%%%%%%%%%%%%%%%%%%%%%%%%%%%%%%%%%%%%%%%%%%%%%%%%%%%%%%%%%%%%%%%%

\section{Introduction}

\label{sec:Intro}
%%%%%%%%%%%%%%%%%%%%%%%%%%%%%%%%%%%%%%%%%%%%%%%%%%%%%%%%%%%%%%%%%%%%%%%%%%%%%

Recently, the LHCb collaboration discovered new resonances $%
T_{cs0}^{a}(2900)^{0/++}$ (in what follows, $T_{cs0}^{a0/++}$) in the
processes $B^{0}\rightarrow \overline{D}^{0}D_{s}^{+}\pi ^{-}$ and $%
B^{+}\rightarrow D^{-}D_{s}^{+}\pi ^{+}$ \cite{LHCb:2022xob,LHCb:2022bkt},
respectively. They were fixed in the $D_{s}^{+}\pi ^{-}$ and $D_{s}^{+}\pi
^{+}$ mass distributions, and are structures with spin-parity $J^{\mathrm{P}%
}=0^{+}$. From the analysis of the decay channels of $T_{cs0}^{a0/++}$ it
becomes clear that they are fully open flavor four-quark systems of $cd%
\overline{s}\overline{u}$/$cu\overline{s}\overline{d}$. Their resonant
parameters are consistent with each other, which means that they are members
of an isospin triplet: This is the first observation of an isospin triplet of
exotic mesons with four different quark flavors. The resonance $%
T_{cs0}^{a++} $ has an additional attractive feature as the first doubly
charged exotic meson discovered experimentally.

It should be emphasized that $T_{cs0}^{a0/++}$ are not first fully open
flavor resonances seen by the LHCb experiment. Indeed, previously LHCb
informed about scalar $X_{0}(2900)$ and vector $X_{1}(2900)$ structures
(hereafter $X_{0}$ and $X_{1}$, respectively), which were found in the $%
D^{-}K^{+}$ invariant mass distribution of the decay $B^{+}\rightarrow
D^{+}D^{-}K^{+}$ \cite{LHCb:2020A,LHCb:2020}. In a four-quark picture both $%
X_{0}$ and $X_{1}$ have the same contents $ud\overline{s}\overline{c}$. New
resonances $T_{cs0}^{a0/++}$ fill up the list of such particles.

The exotic mesons built of four different quarks have already attracted the  interest
of researches. Relevant activities started from announcement by the $D0$
collaboration about the resonance $X(5568)$ \cite{D0:2016mwd,D0:2017qqm}
presumably composed of quarks $su\overline{b}\overline{d}$. Despite the fact
that LHCb, CMS and ATLAS experiments did not confirm existence of this
state, technical tools elaborated during this activity led to some
interesting results, and are still in use in numerous research works. One of
such results is prediction of a charmed partner $X_{c}=[su][\overline{c}%
\overline{d}]$ of $X(5568)$ in the diquark-antidiquark model \ \cite%
{Agaev:2016lkl,Chen:2016mqt}. In our paper \cite{Agaev:2016lkl} it was
investigated in a rather detailed form. Thus, we calculated the mass and
full width of this tetraquark in the context of QCD sum rule method using
different interpolating currents. In the case of scalar-scalar current we
obtained $m_{\mathrm{S}}=(2634\pm 62)~\mathrm{MeV}$ and$\ \Gamma _{\mathrm{S}%
}=(57.7\pm 11.6)~\mathrm{MeV}$, whereas the axial-axial current led to
predictions $m_{\mathrm{A}}=(2590\pm 60)~\mathrm{MeV}$ and $\Gamma _{\mathrm{%
A}}=(63.4\pm 14.2)~\mathrm{MeV}$. It is worth noting that an estimation $%
(2.55\pm 0.09)~\mathrm{GeV}$ for the mass of $X_{c}$ was made in Ref.\ \cite%
{Chen:2016mqt}, as well.

The doubly charged exotic mesons were also objects of interesting analyses.
The $-2|e|$ charged scalar, pseudoscalar and axial-vector
diquark-antidiquarks $Z_{\overline{c}s}=[sd][\overline{u}\overline{c}]$ were
explored in Ref.\ \cite{Agaev:2017oay}. Another class of tetraquarks $%
Z^{++}=[cu][\overline{s}\overline{d}]$ with the electric charge $2|e|$ are
antiparticles of the states $Z_{\overline{c}s}$ and have the same masses and
decay widths. Parameters of the vector tetraquark $Z_{\mathrm{V}}^{++}$ from
this group of particles were found in Ref.\ \cite{Agaev:2021jsz}.

The discovery of the resonances $X_{0}$ and $X_{1}$ highly intensified
investigations of fully open flavor structures \cite%
{Karliner:2020vsi,Wang:2020xyc,Chen:2020aos,Liu:2020nil,Liu:2020orv,Molina:2020hde,Hu:2020mxp,He:2020jna, Lu:2020qmp,Zhang:2020oze,Huang:2020ptc,Xue:2020vtq,Yang:2021izl,Wu:2020job,Abreu:2020ony,Wang:2020prk,Xiao:2020ltm,Dong:2020rgs,Burns:2020xne,Bondar:2020eoa, Chen:2020eyu,Albuquerque:2020ugi,Wang:2021lwy}%
. In these articles different models were suggested to explain the observed
parameters of these states and understand their inner organizations.
Traditionally they were explored in the diquark-antidiquark and hadronic
molecule pictures, which are dominant models to account for similar
experimental data. Thus, $X_{0}$ was treated as a scalar diquark-antidiquark
state $[sc][\overline{u}\overline{d}]$ in Refs.\ \cite%
{Karliner:2020vsi,Wang:2020xyc}. The $X_{0}$ was assigned to be the $S$-wave
hadronic molecule $D^{\ast -}K^{\ast +}$, whereas $X_{1}$ was examined as the $P$%
-wave diquark-antidiquark state $[ud][\overline{c}\overline{s}]$ in Ref.\
\cite{Chen:2020aos}. There were attempts to consider these structures as
rescattering effects. In fact, in Ref.\ \cite{Liu:2020orv} it was asserted
that two resonance-like peaks in the process $B^{+}\rightarrow
D^{+}D^{-}K^{+}$ may be generated by rescattering effects and occur in the
LHCb experiment as the states\ $X_{0}$ and $X_{1}$.

The structures $X_{0}$ and $X_{1}$ were studied in our publications as well.
The mass and width of the resonance $X_{0}$ were calculated in Ref.\ \cite%
{Agaev:2020nrc} in the framework of a hadronic molecule model $\overline{D}%
^{\ast 0}K^{\ast 0}$. Results found in this work for the parameters of $%
X_{0} $ allowed us to confirm its molecule nature. We explored also the
resonance $X_{1}$ by considering it as a vector diquark-antidiquark state $%
X_{\mathrm{V}}=[ud][\overline{c}\overline{s}]$ \cite{Agaev:2021knl}: It
turned out that the diquark-antidiquark structure is an appropriate model to
explain the measured parameters of $X_{1}$.

The LHCb observed only the vector tetraquark $X_{\mathrm{V}}=[ud][\overline{c%
}\overline{s}]$, which was interpreted as $X_{1}$. It is quite possible
that, in near future, the diquark-antidiquark structures $[ud][\overline{c}%
\overline{s}] $ with other quantum numbers will be seen in various exclusive
processes. Therefore, parameters of these yet hypothetical exotic mesons are
necessary to form a theoretical basis for upcoming experimental activities.
Motivated by this reason, we computed the masses and full widths of the
ground-state and radially excited scalar particles $X_{0}^{(\prime )}=[ud][%
\overline{c}\overline{s}]$ \cite{Agaev:2022eeh}. The axial-vector and
pseudoscalar tetraquarks $X_{\mathrm{AV}}$ and $X_{\mathrm{PS}}$ were
investigated in Ref.\ \cite{Sundu:2022kyd}, in which we evaluated their
spectroscopic parameters, i.e., their masses and current couplings, and
widths.

The resonances $T_{cs0}^{a0/++}$ are last experimentally confirmed members
of fully open flavor tetraquark family. In this article we are going to
study the doubly charged state $T_{cs0}^{a++}$, therefore, below, write down
its parameters measured by LHCb \cite{LHCb:2022bkt}:
\begin{eqnarray}
M_{\mathrm{exp}} &=&(2921\pm 17\pm 20)~\mathrm{MeV},  \notag \\
\Gamma _{\mathrm{exp}} &=&(137\pm 32\pm 17)~\mathrm{MeV}.  \label{eq:data1}
\end{eqnarray}

Observation of new tetraquarks $T_{cs0}^{a0/++}$ generated theoretical
investigations aimed to bring them under one of existing models of
four-quark mesons. In our article \cite{Agaev:2022duz}, we argued that
diquark-antidiquark structures are not suitable for these resonances,
because parameters of such states were already evaluated and predictions
obtained for their masses are well below the LHCb data. One of possible ways
to explain $T_{cs0}^{a0/++}$ is to treat them as hadronic molecules. Then $%
T_{cs0}^{a++}$ may be interpreted as a hadronic molecule $D_{s}^{\ast +}\rho
^{+}$ or $D^{\ast +}K^{\ast +}$. In Ref.\ \cite{Agaev:2022duz}, we realized
first of these scenarios, and estimated the mass of molecule $D_{s}^{\ast
+}\rho ^{+}$ by employing the QCD two-point sum rule approach. Our result $%
m=(2917\pm 135)~\mathrm{MeV}$ for the mass of the molecule $D_{s}^{\ast
+}\rho ^{+}$ is consistent with Eq.\ (\ref{eq:data1}).

The resonances $T_{cs0}^{a0/++}$ were investigated in Refs.\ \cite%
{Chen:2022svh,Ge:2022dsp,Wei:2022wtr,Molina:2022jcd,Liu:2022hbk,Yue:2022mnf,Qin:2022nof}
as well, in which authors used different models and calculational schemes.
The one-boson exchange model was employed in Ref.\ \cite{Chen:2022svh} to
study interactions in systems of $D^{(\ast )}K^{(\ast )}$ mesons. Analysis
allowed the authors to assign $T_{cs0}^{a++}$ to be an isovector $D^{\ast
+}K^{\ast +}$ molecule state with the spin-parity $J^{\mathrm{P}}=0^{+}$ and
mass $2891~\mathrm{MeV}$. Interpretation of new tetraquark candidate $%
T_{cs0}^{a}$ as the resonance-like structure generated by threshold effects
was proposed in Ref.\ \cite{Ge:2022dsp}. It was argued that the triangle
singularity induced by the $\chi _{c1}K^{\ast }D^{\ast }$ loop peaks around
the threshold $D^{\ast }K^{\ast }$ and may simulate $T_{cs0}^{a}$. A
multiquark color flux-tube model was used to investigate the resonances $%
T_{cs0}^{a0/++} $ in the context of the diquark-antidiquark model \cite%
{Wei:2022wtr}. The authors found that a system $[cu][\overline{s}\overline{d}%
]$ built of color antitriplet diquark and triplet antidiquark with the mass $%
2923~\mathrm{MeV}$ is a very nice candidate to the resonance $T_{cs0}^{a++}$%
. Features of the charmed-strange tetraquarks were explored also in Ref.\
\cite{Liu:2022hbk} in a nonrelativistic potential quark model. Decays of the
neutral state $T_{cs0}^{a0}$ in the molecular picture were considered in
Ref.\ \cite{Yue:2022mnf}, whereas production mechanisms of the hidden- and
open-charm tetraquarks in $B$ decays were addressed in Ref.\ \cite%
{Qin:2022nof}.

In the present work, we explore the resonance $T_{cs0}^{a++}$ in the context
of the hadronic molecule model. We implement the second scenario and model $%
T_{cs0}^{a++}$ as the hadronic molecule $M=D^{\ast +}K^{\ast +}$. Our
analyses of $M$ include calculations of its mass $m$, current coupling $f$
and full width $\Gamma $. The spectroscopic parameters of $M$ are extracted
from the QCD two-point sum rule computations \cite%
{Shifman:1978bx,Shifman:1978by} by taking into account vacuum expectation
values of different quark, gluon and mixed operators up to dimension $10$.

To estimate the full width of the molecule $M$, we consider its decays to
pairs of conventional mesons $D_{s}^{+}\pi ^{+}$, $D_{s}^{\ast +}\rho ^{+}$,
and $D^{\ast +}K^{\ast +}$. Partial widths of these processes depend on
parameters of initial and final-state particles, as well as on couplings $%
g_{i}$, which determine strong interactions of the molecule $M$ and mesons
at the vertices $MD_{s}^{+}\pi ^{+}$, $MD_{s}^{\ast +}\rho ^{+}$, and $%
MD^{\ast +}K^{\ast +}$, respectively. Because masses and decay constants of
ordinary mesons are known, and spectroscopic parameters of $M$ are object of
present studies, only missed quantities are strong couplings $g_{i}$. The
couplings $g_{i}$ are evaluated in the framework of QCD light-cone sum rule
(LCSR) method \cite{Balitsky:1989ry}. Additionally, to treat technical
peculiarities of four-quark molecule-two conventional meson vertices, we
invoke the technical methods of soft-meson approximation \cite%
{Ioffe:1983ju,Belyaev:1994zk}.

This article is organized in the following way: In Sec.\ \ref{sec:MassCoupl}%
, we find the sum rules for the mass $m$ and current coupling $f$ of the
molecule $M=D^{\ast +}K^{\ast +}$ in the framework of QCD sum rule method.
Numerical analysis of the quantities $m$ and $f$ is carried out in this
section as well, where their values are evaluated. In section \ref%
{sec:Decays}, we investigate the vertices $MD_{s}^{+}\pi ^{+}$, $%
MD_{s}^{\ast +}\rho ^{+}$, and $MD^{\ast +}K^{\ast +}$ and calculate the
corresponding strong couplings $g_{i},\ i=1,2,3$. Obtained information on $%
g_{i}$ is used to find partial widths of these decay channels, and estimate
full width of the molecule $M$. The section \ref{sec:Conclusions} is
reserved for our conclusions.

%%%%%%%%%%%%%%%%%%%%%%%%%%%%%%%%%%%%%%%%%%%%%%%%%%%%%%%%%%%%%%%%%%%%%%%%%%%%%%%%%%%

\section{Mass $m$ and current coupling $f$ of the hadronic molecule $%
M=D^{\ast +}K^{\ast +}$}

\label{sec:MassCoupl}
%%%%%%%%%%%%%%%%%%%%%%%%%%%%%%%%%%%%%%%%%%%%%%%%%%%%%%%%%%%%%%%%%%%%%%%%%%%%%%%%%%%

The key quantity necessary to investigate the spectroscopic parameters of
the molecule $M$ using the QCD two-point sum rule method, is the
interpolating current $J(x)$ for this state. An analytic form of $J(x)$
depends on the structure and constituents of a four-quark exotic meson $%
\overline{q}_{1}\overline{q}_{2}q_{3}q_{4}$. In the molecule picture the
color-singlet structures come from $\mathbf{[1}_{c}\mathbf{]}_{\overline{q}%
_{1}q_{3}}\mathbf{\otimes \lbrack 1}_{c}\mathbf{]}_{\overline{q}_{2}q_{4}}$
and $\mathbf{[8}_{c}\mathbf{]}_{\overline{q}_{1}q_{3}}\mathbf{\otimes
\lbrack 8}_{c}\mathbf{]}_{\overline{q}_{2}q_{4}}$ and ($q_{3}\leftrightarrow
q_{4}$) terms of the color group $SU_{c}(3)$. In the case under discussion,
we assume that the hadronic molecule $M$ is composed of two ordinary vector
mesons $D^{\ast +}$ and $K^{\ast +}$, and restrict our analysis by
singlet-singlet type current. Then, in the $\mathbf{[1}_{c}\mathbf{]}_{%
\overline{d}c}\mathbf{\otimes \lbrack 1}_{c}\mathbf{]}_{\overline{s}u}$
representation $J(x)$ takes the following form
\begin{equation}
J(x)=[\overline{d}_{a}(x)\gamma ^{\mu }c_{a}(x)][\overline{s}_{b}(x)\gamma
_{\mu }u_{b}(x)],  \label{eq:CR1}
\end{equation}%
where $a$ and $b$ are color indices.

This current has the meson-meson structure and is a local product of two
vector currents corresponding to the mesons $D^{\ast +}$ and $K^{\ast +}$.
It couples well to the molecule state $D^{\ast +}K^{\ast +}$. But, at the
same time, $J(x)$ couples also to diquark-antidiquark states, because using
Fierz transformation a molecule current can be presented as a weighted sum
of different diquark-antidiquark currents \cite{Wang:2020rcx}. In its turn,
a diquark-antidiquark current is expressible via molecule structures (for
instance, see Refs.\ \cite{Chen:2022sbf,Xin:2021wcr}). For example, the
current $J(x)$ rewritten in the form
\begin{equation}
J(x)=\delta _{am}\delta _{bn}[\overline{d}_{a}\gamma ^{\mu }c_{m}][\overline{%
s}_{b}\gamma _{\mu }u_{n}],
\end{equation}%
after Fierz transformation contains the vector-vector component%
\begin{equation}
J_{\mathrm{VV}}=-\frac{1}{2}\delta _{am}\delta _{bn}[\overline{d}_{a}\gamma
^{\mu }u_{n}][\overline{s}_{b}\gamma _{\mu }c_{m}].  \label{eq:Fierz1}
\end{equation}%
Using the rearrangement of the color indices $\delta _{am}\delta
_{bn}=\delta _{an}\delta _{bm}+\epsilon _{abk}\epsilon _{mnk},$ with $%
\epsilon _{ijk}$ being the Levi-Civita epsilon, it is not difficult to see
that
\begin{equation}
J_{\mathrm{VV}}=-\frac{1}{2}[\overline{d}_{a}\gamma ^{\mu }u_{a}][\overline{s%
}_{b}\gamma _{\mu }c_{b}]+\cdots .  \label{eq:Fierz2}
\end{equation}%
The term in Eq.\ (\ref{eq:Fierz2}) is the $-D_{s}^{\ast +}\rho ^{+}/2$
meson-meson current, whereas dots indicate the second component which should
be further manipulated to become diquark type current(s). In other words, $%
J(x)$ couples also to a molecule $D_{s}^{\ast +}\rho ^{+}$ and can couple to
other meson pairs such as $D_{s}^{+}\pi ^{+}$ with the same contents and
quantum numbers. Nevertheless, the current $J(x)$ corresponds mainly to the
molecule $D^{\ast +}K^{\ast +}$, which will be demonstrated quantitatively
in the next section by comparing its strong couplings $g_{i}$ to different
two-meson states.

The sum rules for the mass $m$ and current coupling $f$ of the hadronic
molecule $M$ can be obtained from analysis of the correlation function \cite%
{Shifman:1978bx,Shifman:1978by},
\begin{equation}
\Pi (p)=i\int d^{4}xe^{ipx}\langle 0|\mathcal{T}\{J(x)J^{\dag
}(0)\}|0\rangle ,  \label{eq:CF1}
\end{equation}%
with $\mathcal{T}$ being the time-ordering operator.

To derive the required sum rules, the correlator $\Pi (p)$ should be
expressed using the physical parameters of the molecule $M$, as well as
calculated in terms of the fundamental parameters of QCD in quark-gluon
language. The first expression establishes the physical (phenomenological)
side of the sum rules, for which we get
\begin{equation}
\Pi ^{\mathrm{Phys}}(p)=\frac{\langle 0|J|M\rangle \langle M|J^{\dagger
}|0\rangle }{m^{2}-p^{2}}+\cdots .  \label{eq:Phys1}
\end{equation}%
To obtain $\Pi ^{\mathrm{Phys}}(p)$, we insert a complete set of the
intermediate states with the content and quantum numbers of the state $M$
into Eq.\ (\ref{eq:CF1}), and carry out integration over $x$. In Eq.\ (\ref%
{eq:Phys1}), the contribution of the ground-state particle $M$ is isolated
and shown explicitly, whereas dots denote effects due to higher resonances
and continuum states in the $M$ channel.

For further simplification of $\Pi ^{\mathrm{Phys}}(p)$, it is convenient to
introduce the physical parameters of $M$ by means of the matrix element
\begin{equation}
\langle 0|J|M\rangle =fm.  \label{eq:ME1}
\end{equation}%
Then, we get the final formula for the function $\Pi ^{\mathrm{Phys}}(p)$:
\begin{equation}
\Pi ^{\mathrm{Phys}}(p)=\frac{f^{2}m^{2}}{m^{2}-p^{2}}+\cdots .
\label{eq:Phen2}
\end{equation}%
The r.h.s. of Eq.\ (\ref{eq:Phen2}) contains only a trivial Lorentz
structure, which is the unit matrix $\mathrm{I}$. The function $%
f^{2}m^{2}/(m^{2}-p^{2})$ is the invariant amplitude $\Pi ^{\mathrm{Phys}%
}(p^{2})$ that corresponds to this structure: It will be used in our
following analysis.

The QCD side of the sum rules, $\Pi ^{\mathrm{OPE}}(p)$, has to be
calculated in the operator product expansion ($\mathrm{OPE}$) with some
fixed accuracy. To derive $\Pi ^{\mathrm{OPE}}(p)$, we insert the
interpolating current $J(x) $ into Eq.\ (\ref{eq:CF1}), contract the
corresponding heavy and light quark fields, and write the obtained
expression in terms of the corresponding quark propagators. Having carried
out these manipulations, we get for $\Pi ^{\mathrm{OPE}}(p)$,
\begin{eqnarray}
&&\Pi ^{\mathrm{OPE}}(p)=i\int d^{4}xe^{ipx}\mathrm{Tr}\left[ \gamma _{\mu
}S_{c}^{aa^{\prime }}(x)\gamma _{\nu }S_{d}^{a^{\prime }a}(-x)\right]  \notag
\\
&&\times \mathrm{Tr}\left[ \gamma ^{\mu }S_{u}^{bb^{\prime }}(x)\gamma ^{\nu
}S_{s}^{b^{\prime }b}(-x)\right] ,  \label{eq:QCD1}
\end{eqnarray}%
where $S_{c}(x)$ and $S_{u(s,d)}(x)$ are the quark propagators. The explicit
expressions of the heavy and light quarks propagators can be found in Ref.\
\cite{Agaev:2020zad}.

The correlator $\Pi ^{\mathrm{OPE}}(p)$ has also a simple structure $\sim
\mathrm{I}$ and is characterized by an amplitude $\Pi ^{\mathrm{OPE}}(p^{2})$%
. To find a preliminary sum rule, we equate the amplitudes $\Pi ^{\mathrm{%
Phys}}(p^{2})$ and $\Pi ^{\mathrm{OPE}}(p^{2})$. This equality contains
contributions coming from both the ground-state particle and higher
resonances. The latter can be suppressed by applying the Borel
transformation to both sides of the sum rule equality. Afterward, using the
quark-hadron duality assumption, we subtract the suppressed terms from the
obtained expression. These manipulations lead to dependence of the sum rule
equality on the Borel and continuum subtraction (threshold) parameters $%
M^{2} $ and $s_{0}$. Obtained by this way, the expression and its derivative
over $d/d(-1/M^{2})$ allow us to find the sum rules for the mass $m$ and
coupling $f$ of the molecule $M$, which read
\begin{equation}
m^{2}=\frac{\Pi ^{\prime }(M^{2},s_{0})}{\Pi (M^{2},s_{0})},  \label{eq:Mass}
\end{equation}%
and
\begin{equation}
f^{2}=\frac{e^{m^{2}/M^{2}}}{m^{2}}\Pi (M^{2},s_{0}).  \label{eq:Coupl}
\end{equation}%
The function $\Pi (M^{2},s_{0})$ in Eqs.\ (\ref{eq:Mass}) and (\ref{eq:Coupl}%
) is the invariant amplitude $\Pi ^{\mathrm{OPE}}(p^{2})$ after the Borel
transformation and continuum subtraction, and $\Pi ^{\prime
}(M^{2},s_{0})=d\Pi (M^{2},s_{0})/d(-1/M^{2})$.

The Borel transform of the amplitude $\Pi ^{\mathrm{Phys}}(p^{2})$ is given
by the formula
\begin{equation}
\mathcal{B}\Pi ^{\mathrm{Phys}}(p^{2})=fme^{-m^{2}/M^{2}}.
\end{equation}%
For the correlator $\Pi (M^{2},s_{0})$, we find a more complicated expression%
\begin{equation}
\Pi (M^{2},s_{0})=\int_{\mathcal{M}^{2}}^{s_{0}}ds\rho ^{\mathrm{OPE}%
}(s)e^{-s/M^{2}}+\Pi (M^{2}),  \label{eq:InvAmp}
\end{equation}%
where $\mathcal{M}=m_{c}+m_{s}$ is the mass of constituent quarks in the
molecule $M$. It is worth noting, that we neglect masses of $u$ and $d$
quarks, but take into account terms $\sim m_{s}$. At the same time, we do
not include into analysis contributions $\sim m_{s}^{2}$ and set $%
m_{s}^{2}=0 $. The spectral density $\rho ^{\mathrm{OPE}}(s)$ is computed as
an imaginary part of the amplitude $\Pi ^{\mathrm{OPE}}(p^{2})$. Borel
transforms some of terms obtained directly from $\Pi ^{\mathrm{OPE}}(p)$ are
denoted in Eq.\ (\ref{eq:InvAmp}) by $\Pi (M^{2}).$ In this paper,
calculations are carried out by taking into account vacuum condensates up to
dimension $10$. Analytical expressions of $\rho ^{\mathrm{OPE}}(s)$ and $\Pi
(M^{2})$ are lengthy, therefore we do not write down them here explicitly.

For numerical computations of $m$ and $f$, one should specify different
vacuum condensates, which enter to the sum rules in Eqs.\ (\ref{eq:Mass})
and (\ref{eq:Coupl}). These condensates are universal parameters, which were
extracted from analysis of numerous processes: Their numerical values are
listed below:
\begin{eqnarray}
&&\langle \overline{q}q\rangle =-(0.24\pm 0.01)^{3}~\mathrm{GeV}^{3},\
\langle \overline{s}s\rangle =(0.8\pm 0.1)\langle \overline{q}q\rangle ,
\notag \\
&&\langle \overline{q}g_{s}\sigma Gq\rangle =m_{0}^{2}\langle \overline{q}%
q\rangle ,\ \langle \overline{s}g_{s}\sigma Gs\rangle =m_{0}^{2}\langle
\overline{s}s\rangle ,  \notag \\
&&m_{0}^{2}=(0.8\pm 0.2)~\mathrm{GeV}^{2}  \notag \\
&&\langle \frac{\alpha _{s}G^{2}}{\pi }\rangle =(0.012\pm 0.004)~\mathrm{GeV}%
^{4},  \notag \\
&&\langle g_{s}^{3}G^{3}\rangle =(0.57\pm 0.29)~\mathrm{GeV}^{6},  \notag \\
&&m_{s}=93_{-5}^{+11}~\mathrm{MeV},\ m_{c}=1.27\pm 0.02~\mathrm{GeV}.
\label{eq:Parameters}
\end{eqnarray}%
We have also  included into Eq.\ (\ref{eq:Parameters})  the masses of $c$ and $%
s$ quarks.

\begin{figure}[h]
\includegraphics[width=8.5cm]{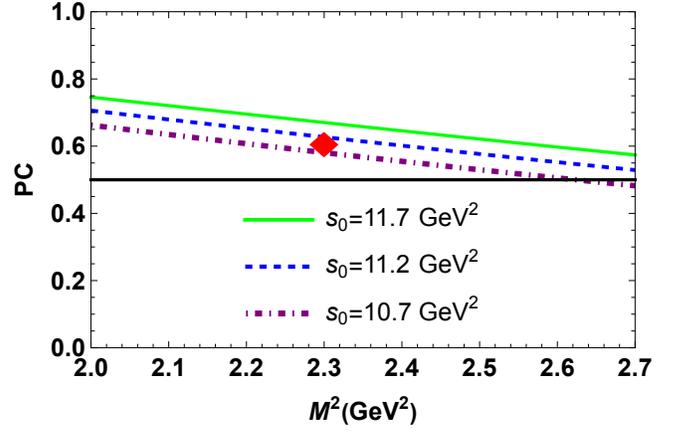}
\caption{Dependence of the pole contribution on the Borel parameter $M^{2}$
at fixed $s_{0}$. The limit $\mathrm{PC}=0.5$ is shown by the horizontal
black line. The red diamond notes the point, where the mass $m$ of the
molecule $M=D^{\ast +}K^{\ast +}$ has effectively been calculated. }
\label{fig:PC}
\end{figure}

The sum rules in Eqs.\ (\ref{eq:Mass}) and (\ref{eq:Coupl}) depend also on
the Borel and continuum threshold parameters $M^{2}$ and $s_{0}$. The choice
of working windows for $M^{2}$ and $s_{0}$ has to satisfy standard
constraints imposed on the pole contribution ($\mathrm{PC}$) and convergence
of the operator product expansion. To quantify these constraints, it is
appropriate to use expressions
\begin{equation}
\mathrm{PC}=\frac{\Pi (M^{2},s_{0})}{\Pi (M^{2},\infty )},  \label{eqPC}
\end{equation}%
and
\begin{equation}
R(M^{2})=\frac{\Pi ^{\mathrm{DimN}}(M^{2},s_{0})}{\Pi (M^{2},s_{0})},
\label{eq:Convergence}
\end{equation}%
where $\Pi ^{\mathrm{DimN}}(M^{2},s_{0})=$ $\Pi ^{\mathrm{Dim(8+9+10)}%
}(M^{2},s_{0})$.

The $\mathrm{PC}$ is used to fix maximum of the Borel parameter $M_{\mathrm{%
max}}^{2}$, whereas its minimal value $M_{\mathrm{min}}^{2}$ is limited by
the convergence of $\mathrm{OPE}$. In sum rule analyses of ordinary hadrons $%
\mathrm{PC}\geq 0.5$ is a standard requirement. When studying multiquark
particles this constraint reduces the region of the allowed $M^{2}$. To be
convinced in convergence of the operator product expansion, we demand
fulfillment of $R(M_{\mathrm{min}}^{2})\leq 0.05$.

Our calculations prove that the regions for the parameters $M^{2}$ and $%
s_{0} $,
\begin{equation}
M^{2}\in \lbrack 2,2.7]~\mathrm{GeV}^{2},\ s_{0}\in \lbrack 10.7,11.7]~%
\mathrm{GeV}^{2},  \label{eq:Wind1a}
\end{equation}%
meet all required restrictions. Thus, at $M^{2}=2.7~\mathrm{GeV}^{2}$ the
pole contribution on average in $s_{0}$ is $0.53$, whereas at $M^{2}=2~%
\mathrm{GeV}^{2}$ it equals to $0.72$. In Fig.\ \ref{fig:PC} the pole
contribution is plotted as a function of $M^{2}$ at various fixed $s_{0}$.
It is seen that by excluding the small region $M^{2}\geq 2.65~\mathrm{GeV}%
^{2}$ at $s_{0}=10.7~\mathrm{GeV}^{2}$ the pole contribution exceeds $0.5$.
On average in $s_{0}$, the constraint $\mathrm{PC}\geq 0.5$ is satisfied in
the all working window for the Borel parameter. At the minimum point $%
M^{2}=2~\mathrm{GeV}^{2}$, we find $R(2~\mathrm{GeV}^{2})\approx 0.015$ and
the sum of dimension-$8,9,10$ contributions is less than $1.5\%$ of the full
result.

Dominance of the perturbative contribution to $\Pi (M^{2},s_{0})$, and
convergence of the operator product expansion are another important problems
in the sum rule studies. In Fig.\ \ref{fig:Conv}, we compare the
perturbative and nonperturbative components of the correlation function. One
sees that the perturbative contribution to $\Pi (M^{2},s_{0})$ prevails over
nonperturbaative one, and forms more than $53\%$ of $\Pi (M^{2},s_{0})$
already at $M^{2}=2~\mathrm{GeV}^{2}$ growing gradually in the considered
range of $M^{2}$. From this figure it is also clear that convergence of $%
\mathrm{OPE}$ is satisfied: Contributions of the nonperturbative terms
reduce by increasing the dimensions of the corresponding operators. There is
some disordering in these contributions connected with smallness of gluon
condensates. The dimension-$3,\ 6,\ 9$ and $10$ terms are positive. The $%
\mathrm{Dim3}$ and $\mathrm{Dim6}$ contributions numerically exceed
contributions of other operators, whereas $\mathrm{Dim9}$ and $\mathrm{Dim10}
$ terms are very small and not shown in the plot.

Predictions for $m$ and $f$ are obtained by taking the mean values of these
parameters calculated at different choices of $M^{2}$ and $s_{0}$:
\begin{eqnarray}
m &=&(2924~\pm 107)~\mathrm{MeV},  \notag \\
f &=&(4.3\pm 0.7)\times 10^{-3}~\mathrm{GeV}^{4}.  \label{eq:Result1}
\end{eqnarray}%
The $m$ and $f$ from Eq.\ (\ref{eq:Result1}) effectively correspond to the
sum rules' results at $M^{2}=2.3~\mathrm{GeV}^{2}$ and $s_{0}=11.2~\mathrm{%
GeV}^{2}$ noted in Fig.\ \ref{fig:PC} by the red diamond. This point is
approximately at the middle of the intervals shown in Eq.\ (\ref{eq:Wind1a}%
), where the pole contribution is $\mathrm{PC}\approx 0.62$. The
circumstances discussed above ensure the ground-state nature of $M$ and
reliability of the obtained results.

The dependence of the mass $m$ on the parameters $M^{2}$ and $s_{0}$ is
drawn in Fig.\ \ref{fig:Mass}. In general, a physical quantity should not
depend on the auxiliary parameter of computations $M^{2}$. Nevertheless,
such residual dependence of $m$ on the Borel parameter, as well as on $s_{0}$
exists and generates theoretical ambiguities of the extracted predictions in
Eq.\ (\ref{eq:Result1}). It is worth noting that these ambiguities are
smaller for $m$ than for $f$. Indeed, for the mass they are equal only to $%
\pm 4\%$ of the central value, whereas in the case of $f$, they amount to $%
\pm 16\%$. Such difference is connected by the analytical forms of the sum
rules for these quantities: The mass $m$ is given by the ratio of the
correlation functions which smooths relevant effect, whereas $f$ depends on $%
\Pi (M^{2},s_{0})$ itself.

Our result for the mass of the molecule $D^{\ast +}K^{\ast +}$ agrees very
well with the LHCb datum. This is necessary, but not enough to make credible
conclusions about the nature of the resonance $T_{cs0}^{a++}$: For more
reliable statements, one needs to estimate also the full width of the
hadronic molecule $D^{\ast +}K^{\ast +}$ suggested in this paper to model $%
T_{cs0}^{a++}$.

\begin{widetext}

\begin{figure}[h!]
\begin{center}
\includegraphics[totalheight=6cm,width=8cm]{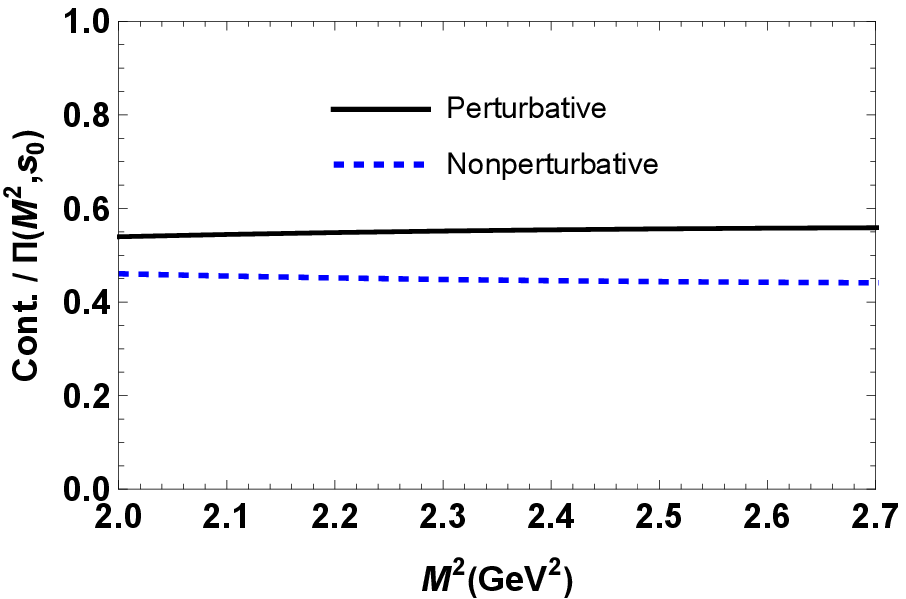}
\includegraphics[totalheight=6cm,width=8cm]{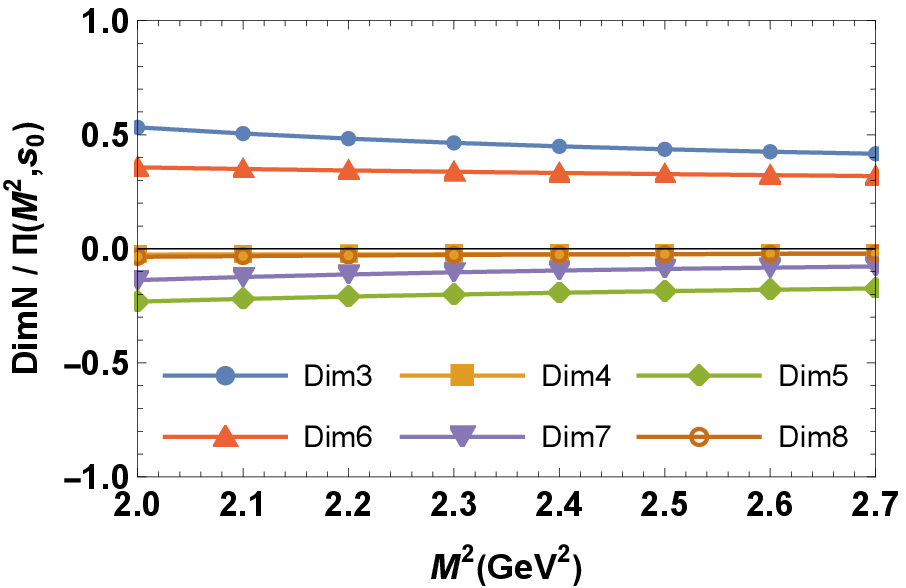}
\end{center}
\caption{Left: Different contributions to $\Pi
(M^{2},s_{0})$ normalized to $1$ as functions of the Borel parameter $M^2$,
Right: Normalized contributions of various operators to $\Pi
(M^{2},s_{0})$ as functions of $M^2$. All curves in this figure have been calculated at $s_0=11.2~\mathrm{GeV}^2$.}
\label{fig:Conv}
\end{figure}

\begin{figure}[h!]
\begin{center}
\includegraphics[totalheight=6cm,width=8cm]{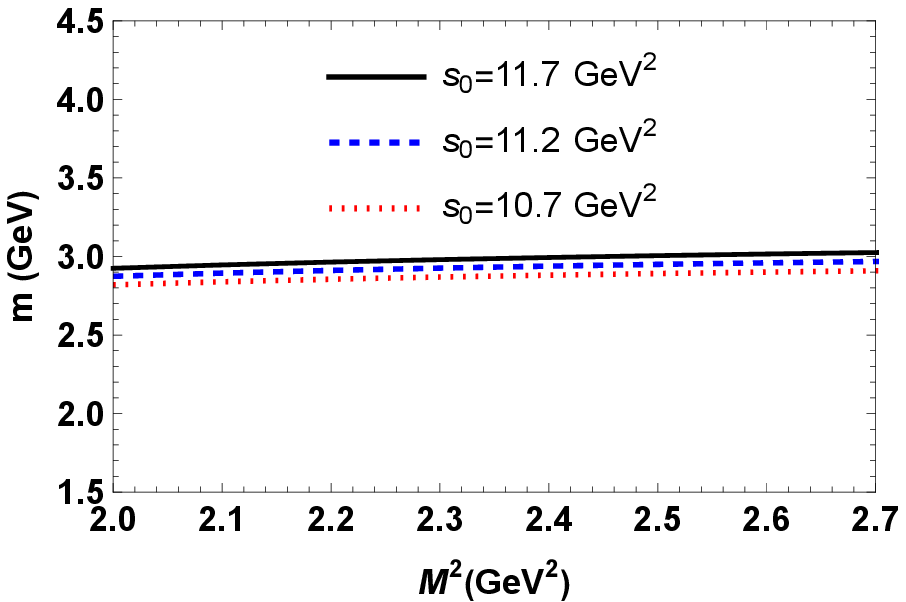}
\includegraphics[totalheight=6cm,width=8cm]{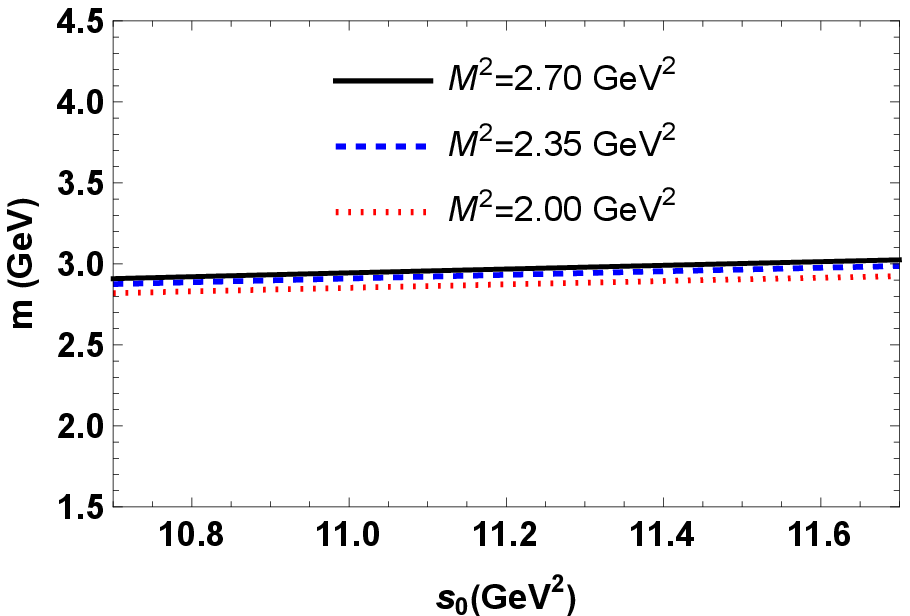}
\end{center}
\caption{Mass $m$ of the molecule $M$ as functions of the Borel $M^{2}$ (left panel), and continuum threshold $s_0$ parameters (right panel).}
\label{fig:Mass}
\end{figure}

\end{widetext}

%%%%%%%%%%%%%%%%%%%%%%%%%%%%%%%%%%%%%%%%%%%%%%%%%%%%%%%%%%%%%%%%%%%%%%%%%%%

\section{Width of the molecule $M$}

\label{sec:Decays}
%%%%%%%%%%%%%%%%%%%%%%%%%%%%%%%%%%%%%%%%%%%%%%%%%%%%%%%%%%%%

The quark content, spin-parity and mass of the molecule $M=$ $D^{\ast
+}K^{\ast +}$ allow us to classify its decay channels. Because $%
T_{cs0}^{a++} $ was seen in the $D_{s}^{+}\pi ^{+}$ invariant mass
distribution, the process $M\rightarrow D_{s}^{+}\pi ^{+}$ should be the
dominant decay channel of the molecule $M$. The processes $M\rightarrow
D_{s}^{\ast +}\rho ^{+}$ and $M\rightarrow D^{\ast +}K^{\ast +}$ are also
among the kinematically allowed decay modes of the $M$. We are going to
evaluate the full width of the molecule $M$ using these decay channels. It
is  noting that, $\rho ^{+}$ and $K^{\ast +}$ mesons decay almost
exclusively to $\pi ^{+}\pi ^{0}$ and $(K\pi )^{+}$ pairs (see, Ref. \cite%
{PDG:2022}), therefore widths of the last two processes can be considered
also as widths of the modes $M\rightarrow D_{s}^{\ast +}\pi ^{+}\pi ^{0}$
and $M\rightarrow D^{\ast +}(K\pi )^{+}$, respectively.

Partial widths of aforementioned processes are determined by the strong
couplings $g_{i}$ at vertices $MD_{s}^{+}\pi ^{+}$, and etc. One of the
effective ways to evaluate them is the QCD light-cone sum rule method \cite%
{Balitsky:1989ry}. In this approach, the QCD side of the sum rule, instead
of the local vacuum condensates, is expressed in terms of one of the final
mesons' distribution amplitudes (DAs). In general, the DAs of a hadron are
nonlocal matrix elements of various operators with different twists
sandwiched between the hadron and vacuum states. They are specific for each
particle and modeled by taking into account the available experimental data.
The LCSR method is suitable for analysis of not only the conventional
hadrons, but also the multiquark systems such as tetraquarks, pentaquarks,
etc. In the case of tetraquark-meson-meson vertices, however, these DAs
reduce to the local matrix elements of the meson. For treatment of such
vertices, one needs to apply some additional mathematical tools. The LCSR
method was adapted for investigation of the tetraquark-meson-meson vertices
in Ref.\ \cite{Agaev:2016dev} and applied to study numerous decays of
four-quark exotic mesons \cite{Agaev:2020zad}.

%%%%%%%%%%%%%%%%%%%%%%%%%%%%%%%%%%%%%%%%%%%%%%%%%%%%%%%%%%%%%%%%%%%%%%%%%%%%%

\subsection{Decay $M\rightarrow D_{s}^{+}\protect\pi ^{+}$}

%%%%%%%%%%%%%%%%%%%%%%%%%%%%%%%%%%%%%%%%%%%%%%%%%%%%%%%%%%%%%%%%%%%%%

Here, we consider, in a detailed manner, the decay of the molecule $M$ to
pair of pseudoscalar mesons $D_{s}^{+}\pi ^{+}$. Partial width of this
process depends on the spectroscopic parameters of the initial and
final-state particles. The spectroscopic parameters of $M$ have been
evaluated in the previous section. The masses and decay constants of the
mesons $D_{s}^{+}$ and $\pi ^{+}$ are available from other sources. The only
unknown quantity required to calculate the $M\rightarrow D_{s}^{+}\pi ^{+}$
decays's partial width is the strong coupling $g_{1}$ of the particles at
the vertex $MD_{s}^{+}\pi ^{+}$.

The coupling $g_{1}$ is defined in terms of the on-mass-shell matrix
element,
\begin{equation}
\langle \pi (q)D_{s}\left( p\right) |M(p^{\prime })\rangle =g_{1}p\cdot
p^{\prime },  \label{eq:Mel2}
\end{equation}%
where the mesons $\pi ^{+}$ and $D_{s}^{+}$ are denoted as $\pi $ and $D_{s}$%
, respectively. In Eq.\ (\ref{eq:Mel2}) $p^{\prime }$, $p$ and $q$ are
four-momenta of $M$, and mesons $D_{s}$ and $\pi $.

In the framework of the LCSR method the sum rule for the coupling $g_{1}$
can be obtained from the correlation function,
\begin{equation}
\Pi (p,q)=i\int d^{4}xe^{ipx}\langle \pi (q)|\mathcal{T}\{J^{D_{s}}(x)J^{%
\dag }(0)\}|0\rangle ,  \label{eq:CorrF3}
\end{equation}%
where $J(x)$ is the current for the molecule $M$ given by Eq.\ (\ref{eq:CR1}%
). The interpolating current $J^{D_{s}}(x)$ for the meson $D_{s}^{+}$ is
defined by the formula
\begin{equation}
J^{D_{s}}(x)=\overline{s}_{j}(x)i\gamma _{5}c_{j}(x),  \label{eq:Dcur}
\end{equation}%
with $j$ being the color index.

At this stage of our analysis, we express the correlation function $\Pi
(p,q) $ using the physical parameters of the particles involved into the
decay process. To this end, we write $\Pi ^{\mathrm{Phys}}(p,q)$ in the
factorized form \cite{Belyaev:1994zk,Agaev:2022iha},
\begin{eqnarray}
\Pi ^{\mathrm{Phys}}(p,q) &=&\frac{\langle \pi (q)D_{s}\left( p\right)
|M(p^{\prime })\rangle \langle M(p^{\prime })|J^{\dagger }|0\rangle }{\left(
p^{\prime 2}-m^{2}\right) }  \notag \\
&&\times \frac{\ \langle 0|J^{D_{s}}|D_{s}(p)\rangle }{\left(
p^{2}-m_{D_{s}}^{2}\right) },  \label{eq:CorrF4}
\end{eqnarray}%
where $m_{D_{s}}$ is the mass of the meson $D_{s}^{+}$. As is seen, the
function $\Pi ^{\mathrm{Phys}}(p,q)$ contains the matrix elements of the
vertex $MD_{s}^{+}\pi ^{+}$, the molecule $M$ and meson $D_{s}^{+}$. The
matrix element of $M$ is known from Eq.\ (\ref{eq:ME1}), whereas for the
meson $D_{s}^{+}$ we use
\begin{equation}
\langle 0|J^{D_{s}}|D_{s}(p)\rangle =\frac{f_{D_{s}}m_{D_{s}}^{2}}{%
m_{c}+m_{s}},
\end{equation}%
with $f_{D_{s}}$ being the decay constant of $D_{s}^{+}$.

After simple manipulations, we get
\begin{eqnarray}
&&\Pi ^{\mathrm{Phys}}(p,q)=g_{1}\frac{fmf_{D_{s}}m_{D_{s}}^{2}}{%
(m_{c}+m_{s})\left( p^{2}-m_{D_{s}}^{2}\right) }  \notag \\
&&\times \frac{1}{\left( p^{\prime 2}-m^{2}\right) }p\cdot p^{\prime
}+\cdots .  \label{eq:CorrF5}
\end{eqnarray}%
The term in Eq.\ (\ref{eq:CorrF5}) corresponds to the contribution of
ground-state particles in $M$ and $D_{s}^{+}$ channels. Effects of higher
resonances and continuum states in these channels are denoted by ellipses.
The function $\Pi ^{\mathrm{Phys}}(p,q)$ form the physical side of the sum
rule for the coupling $g_{1}$. It has the Lorentz structure proportional to
the unit matrix\ $\mathrm{I}$, therefore the expression in r.h.s. of Eq.\ (%
\ref{eq:CorrF5}) is the invariant amplitude $\Pi ^{\mathrm{Phys}%
}(p^{2},p^{\prime 2})$, which depends on two variables $p^{2}$ and $%
p^{\prime 2}$.

The correlation function $\Pi (p,q)$ calculated in terms of quark
propagators and matrix elements of the pion constitutes the QCD side of sum
rules and is equal to
\begin{eqnarray}
&&\Pi ^{\mathrm{OPE}}(p,q)=-\int d^{4}xe^{ipx}\left[ \gamma _{\mu
}S_{s}^{bj}(-x){}\gamma _{5}S_{c}^{ja}(x)\gamma ^{\mu }\right] _{\alpha
\beta }  \notag \\
&&\times \langle \pi (q)|\overline{u}_{\alpha }^{b}(0)d_{\beta
}^{a}(0)|0\rangle ,  \label{eq:CorrF6}
\end{eqnarray}%
where $\alpha $ and $\beta $ are the spinor indices.

The correlator $\Pi ^{\mathrm{OPE}}(p,q)$, apart from propagators of the $c$
and $s$ quarks, contains also the local matrix elements $\langle \pi |%
\overline{u}_{\alpha }^{b}d_{\beta }^{a}|0\rangle $ of the pion. In the
standard LCSR method, while studying vertices of conventional mesons, the
correlator depends on the nonlocal matrix elements of the meson (for
instance, $\pi $ meson), which after some transformations can be expressed
in terms of its different DAs. In the case under discussion, the $\Pi ^{%
\mathrm{OPE}}(p,q)$ contains the pion's local matrix elements appearance of
which has a simple explanation. Indeed, because the hadronic molecule $M$ is
built of four valence quarks located at space-time position $x=0$,
contractions of two quark operators from the currents $J^{D_{s}}(x)$ and $%
J(x)$ leave free the two quark fields from $M$ at the same position $x=0$.
This feature of the correlation function $\Pi ^{\mathrm{OPE}}(p,q)$
connected with differences in the quark contents of tetraquarks and ordinary
mesons is unavoidable effect for all tetraquark-meson-meson vertices.

It turns out that the $\Pi ^{\mathrm{OPE}}(p,q)$-type correlators emerge in
the limit $q\rightarrow 0$ in LCSR calculations \cite{Belyaev:1994zk}, which
is known as the soft-meson approximation. In this approximation, $%
p=p^{\prime }$ and invariant amplitudes $\Pi ^{\mathrm{Phys}}(p^{2})$ and $%
\Pi ^{\mathrm{OPE}}(p^{2})$ depend only on one variable $p^{2}$. It is worth
emphasizing that the limit $q\rightarrow 0$ is applied to hard parts of the
invariant amplitudes, whereas in their soft parts (i.e., in matrix elements)
terms $\sim q^{2}=m^{2}$ are taken into account. In other words, soft-meson
approximation should not be considered as a massless limit of the
correlation functions.

Technical difficulties generated by this limit in the physical side of sum
rules can be cured by means of technical tools elaborated in Refs.\ \cite%
{Ioffe:1983ju,Belyaev:1994zk}. It is important that the sum rules for the
strong couplings obtained using the full LCSR method, and soft-meson
approximation lead to numerically close results \cite{Belyaev:1994zk}. To
clarify this last point, we note that in the full version of the LCSR method
a sum rule for the strong coupling at a vertex of three conventional mesons
depends on numerous two- and three-particle quark-gluon DAs of a final
meson. In the limit $q\rightarrow 0$ in this expression survive only a few
leading terms. Because their contributions are numerically decisive, for the
strong coupling the full version and soft-meson\ limit of the light-cone sum
rules give close predictions. Thus, in Ref.\ \cite{Belyaev:1994zk} the
couplings $g_{D^{\ast }D\pi }$ and $g_{B^{\ast }B\pi }$ at the vertices $%
D^{\ast }D\pi $ and $B^{\ast }B\pi $ were calculated using both of these
methods. In the full LCSR approach these couplings are equal to
\begin{equation}
g_{D^{\ast }D\pi }^{\mathrm{full}}=12.5\pm 1.0,\ g_{B^{\ast }B\pi }^{\mathrm{%
full}}=29\pm 3,  \label{eq:SumR1}
\end{equation}%
whereas in the soft-meson approximation, the authors found
\begin{equation}
g_{D^{\ast }D\pi }^{\mathrm{soft}}=11\pm 2,\ g_{B^{\ast }B\pi }^{\mathrm{soft%
}}=28\pm 6.  \label{eq:SumR2}
\end{equation}%
As is seen, values of the couplings are very close to each other, though
uncertainties in the soft-limit are larger than in the full version \cite%
{Belyaev:1994zk}.

These arguments do not imply a necessity of the soft-meson
approximation to study all of the vertices containing four-quark states. Two
tetraquark-meson vertices, for example, can be readily investigated in the
context of the conventional LCSR method \cite{Agaev:2016srl}.

The local matrix elements, $\langle \pi |\overline{u}_{\alpha }^{b}d_{\beta
}^{a}|0\rangle $, carry color and spinor indices and are uneasy objects for
further operations. We can rewrite them in convenient forms by expanding $%
\overline{u}d$ over the full set of Dirac matrices $\Gamma ^{J}$,
\begin{equation}
\Gamma ^{J}=\mathbf{1},\ \gamma _{5},\ \gamma _{\mu },\ i\gamma _{5}\gamma
_{\mu },\ \sigma _{\mu \nu }/\sqrt{2},  \label{eq:Dirac}
\end{equation}%
and projecting onto the colorless states
\begin{equation}
\overline{u}_{\alpha }^{b}(0)d_{\beta }^{a}(0)\rightarrow \frac{1}{12}\delta
^{ba}\Gamma _{\beta \alpha }^{J}\left[ \overline{u}(0)\Gamma ^{J}d(0)\right]
.  \label{eq:MatEx}
\end{equation}%
The operators $\overline{u}\Gamma ^{J}d$, sandwiched between the $\pi $
meson and vacuum, generate local matrix elements of the $\pi $ meson, which
can be implemented into $\Pi ^{\mathrm{OPE}}(p,q)$.

The expression obtained for the $\Pi ^{\mathrm{OPE}}(p^{2})$ in the soft
limit is considerably more simple than the invariant amplitude in the full
version of LCSR method. But, at the same time, soft-meson approximation
produces problems in the physical side of the sum rule. Below, we will come
back to finish calculation of $\Pi ^{\mathrm{OPE}}(p^{2})$, but now it is
time to fix sources of complications in $\Pi ^{\mathrm{Phys}%
}(p^{2},p^{\prime 2})$. To this end, we rewrite this amplitude in the soft
limit
\begin{eqnarray}
&&\Pi ^{\mathrm{Phys}}(p^{2})=g_{1}\frac{fmf_{D_{s}}m_{D_{s}}^{2}}{%
2(m_{c}+m_{s})\left( p^{2}-\widetilde{m}^{2}\right) ^{2}}  \notag \\
&&\times (2\widetilde{m}^{2}-m_{\pi }^{2})+\cdots ,  \label{eq:CorrF5a}
\end{eqnarray}%
where $\widetilde{m}^{2}=(m^{2}+m_{D_{s}}^{2})/2$ and $m_{\pi }$ is the mass
of the pion. It is seen that the amplitude $\Pi ^{\mathrm{Phys}}(p^{2})$, in
the soft approximation, instead of two poles at different points has one
double pole at $p^{2}=\widetilde{m}^{2}$.

The Borel transformation of $\Pi ^{\mathrm{Phys}}(p^{2})$ is given by the
expression,
\begin{eqnarray}
&&\Pi ^{\mathrm{Phys}}(M^{2})=\left[ g_{1}\frac{fmf_{D_{s}}m_{D_{s}}^{2}}{%
2(m_{c}+m_{s})}(2\widetilde{m}^{2}-m_{\pi }^{2})\right.  \notag \\
&&\left. +AM^{2}\right] \frac{e^{-\widetilde{m}^{2}/M^{2}}}{M^{2}}+\cdots .
\label{eq:CorrF5b}
\end{eqnarray}%
Apart from the ground-state contribution, in the soft limit, the amplitude $%
\Pi ^{\mathrm{Phys}}(M^{2})$ contains additional unsuppressed terms $\sim A$
\cite{Belyaev:1994zk}. These terms correspond to the transitions from the
excited states in $M=D^{\ast +}K^{\ast +}$ channel with $m^{\ast }>m$, and
are not suppressed relative to the ground-state contribution even after the
Borel transformation. These circumstances make problematic extraction of the
$g_{1}$ from Eq.\ (\ref{eq:CorrF5b}). The contributions $\sim A$ can be
removed from the physical side of the sum rule by means of the operator $%
\mathcal{P}(M^{2},\widetilde{m}^{2})$ \cite{Ioffe:1983ju,Belyaev:1994zk},
\begin{equation}
\mathcal{P}(M^{2},\widetilde{m}^{2})=\left( 1-M^{2}\frac{d}{dM^{2}}\right)
M^{2}e^{\widetilde{m}^{2}/M^{2}},  \label{eq:Oper}
\end{equation}%
which should be applied to both sides of the sum rule equality.

After this operation the remaining suppressed terms in $\Pi ^{\mathrm{Phys}%
}(M^{2})$, denoted in Eq.\ (\ref{eq:CorrF5b}) by ellipses, can be subtracted
by a standard way. As a result, we find the sum rule for the strong coupling
$g_{1}$, which reads
\begin{equation}
g_{1}=\frac{2(m_{c}+m_{s})}{fmf_{D_{s}}m_{D_{s}}^{2}(2\widetilde{m}%
^{2}-m_{\pi }^{2})}\mathcal{P}(M^{2},\widetilde{m}^{2})\Pi ^{\mathrm{OPE}%
}(M^{2},s_{0}),  \label{eq:SRcoupl}
\end{equation}%
where $\Pi ^{\mathrm{OPE}}(M^{2},s_{0})$ is the Borel transformed and
subtracted invariant amplitude $\Pi ^{\mathrm{OPE}}(p^{2})$.

Recipes to compute the correlation function $\Pi ^{\mathrm{OPE}}(p,q)$ in
the soft approximation were explained in Ref.\ \cite{Agaev:2016dev},
therefore we give only principal points of these calculations. Thus, having
substituted the expansion (\ref{eq:MatEx}) into Eq.\ (\ref{eq:CorrF6}), we
perform summations over color indices and fix the local matrix elements of
the pion that contribute to the $\Pi ^{\mathrm{OPE}}(p,q)$ in the soft-meson
limit. It turns out that the contributions to $\Pi ^{\mathrm{OPE}}(p,q=0)$
come from the pion's two-particle twist-3 matrix element,
\begin{equation}
\langle 0|\overline{d}i\gamma _{5}u|\pi \rangle =f_{\pi }\mu _{\pi },
\label{eq:MatElK1}
\end{equation}%
where
\begin{equation}
\mu _{\pi }=\frac{m_{\pi }^{2}}{m_{u}+m_{d}}=-\frac{2\langle \overline{q}%
q\rangle }{f_{\pi }^{2}}.  \label{eq:PCAC}
\end{equation}%
The second equality in Eq.\ (\ref{eq:PCAC}) arises from the partial
conservation of axial-vector current (PCAC).

The amplitude $\Pi ^{\mathrm{OPE}}(M^{2},s_{0})$ is given by the formula
\begin{eqnarray}
&&\Pi ^{\mathrm{OPE}}(M^{2},s_{0})=\frac{f_{\pi }\mu _{\pi }}{8\pi ^{2}}%
\int_{\mathcal{M}^{2}}^{s_{0}}\frac{ds(m_{c}^{2}-s)}{s}  \notag \\
&&\times (m_{c}^{2}-2m_{c}m_{s}-s)e^{-s/M^{2}}+f_{\pi }\mu _{\pi }\Pi _{%
\mathrm{NP}}(M^{2}).  \notag \\
&&  \label{eq:DecayCF}
\end{eqnarray}%
The first term in Eq.\ (\ref{eq:DecayCF}) is the perturbative component of
the $\Pi ^{\mathrm{OPE}}(M^{2},s_{0})$. The nonperturbative term $\Pi _{%
\mathrm{NP}}(M^{2})$ is computed with dimension-$9$ accuracy and has the
following form:%
\begin{eqnarray}
&&\Pi _{\mathrm{NP}}(M^{2})=\frac{\langle \overline{s}s\rangle }{6M^{2}}%
\left[ M^{2}(2m_{c}-m_{s})-m_{c}^{2}m_{s}\right] e^{-m_{c}^{2}/M^{2}}  \notag
\\
&&+\langle \frac{\alpha _{s}G^{2}}{\pi }\rangle \frac{m_{c}}{72M^{4}}%
\int_{0}^{1}\frac{dx}{(1-x)^{3}}\left[ m_{s}(1-x)(2M^{2}+m_{c}^{2})\right.
\notag \\
&&\left. -m_{c}^{3}x\right] e^{-m_{c}^{2}/[M^{2}(1-x)]}-\frac{\langle
\overline{s}g\sigma Gs\rangle m_{c}^{3}}{6M^{6}}(M^{2}-m_{c}m_{s})  \notag \\
&&\times e^{-m_{c}^{2}/M^{2}}+\langle \frac{\alpha _{s}G^{2}}{\pi }\rangle
\langle \overline{s}s\rangle \frac{\pi ^{2}m_{c}}{9M^{8}}\left[
M^{2}m_{c}(m_{c}+m_{s})\right.  \notag \\
&&\left. -2m_{c}^{3}m_{s}-M^{4}\right] e^{-m_{c}^{2}/M^{2}}+\langle \frac{%
\alpha _{s}G^{2}}{\pi }\rangle \langle \overline{s}g\sigma Gs\rangle \frac{%
\pi ^{2}m_{c}}{18M^{12}}  \notag \\
&&\times \left[ 3M^{6}-20m_{c}^{5}m_{s}+10m_{c}^{3}M^{2}(m_{c}+2m_{s})\right.
\notag \\
&&\left. -4M^{4}m_{c}(3m_{c}+m_{s})\right] e^{-m_{c}^{2}/M^{2}}.
\end{eqnarray}

\begin{table}[tbp]
\begin{tabular}{|c|c|}
\hline\hline
Parameters & Values (in $\mathrm{MeV}$ units) \\ \hline
$m_{D_s}$ & $1969.0\pm 1.4$ \\
$m_{\pi}$ & $139.57039 \pm 0.00017$ \\
$m_{D_s^{\ast}}$ & $2112.2\pm 0.4$ \\
$m_{\rho}$ & $775.11\pm 0.34$ \\
$m_{D^{\ast}}$ & $2010.26\pm 0.05$ \\
$m_{K^{\ast}}$ & $891.67\pm 0.26$ \\
$f_{D_s}$ & $249.9 \pm 0.5$ \\
$f_{\pi}$ & $130.2 \pm 0.8$ \\
$f_{D_s^{\ast}}$ & $268.8\pm 6.5$ \\
$f_{\rho}$ & $216 \pm 3$ \\
$f_{D^{\ast}}$ & $223.5 \pm 0.5$ \\
$f_{K^{\ast}}$ & $204\pm 0.3$ \\ \hline\hline
\end{tabular}%
\caption{Masses and decay constants of the mesons $D_s^{(\ast)}$, $D^{\ast} $%
, $K^{\ast}$, $\protect\pi$ and $\protect\rho$, which have been used in
numerical computations. }
\label{tab:Param}
\end{table}

Besides the vacuum condensates, the sum rule in Eq.\ (\ref{eq:SRcoupl})
contains also the masses and decay constants of the final-state mesons $%
D_{s}^{+}$ and $\pi ^{+}$. Numerical values of these parameters, as well as
parameters of other mesons, which will be necessary later, are presented in
Table\ \ref{tab:Param}. For the decay constants of the mesons $D_{s}^{\ast
+} $ and $D^{\ast +}$, we employ predictions of the QCD lattice method \cite%
{Lubicz:2016bbi}. For all other parameters, we use information of the
Particle Data Group, mainly its last edition Ref.\ \cite{PDG:2022}. All
these input parameters were either measured experimentally or extracted by
means of alternative theoretical approaches.

Numerical analysis demonstrates that the working regions shown in Eq.\ (\ref%
{eq:Wind1a}) used in calculations of the $M$ molecule's mass meet the
required restrictions on the Borel and continuum subtraction parameters $%
M^{2}$ and $s_{0}$ imposed in the case of the decay process. Therefore, in
computations of $\Pi ^{\mathrm{OPE}}(M^{2},s_{0})$ the parameters $M^{2}$
and $s_{0}$ have been varied within the limits (\ref{eq:Wind1a}).

For $g_{1}$, the numerical calculations yield
\begin{equation}
g_{1}=(7.1\pm 1.1)\times 10^{-1}~\mathrm{GeV}^{-1}.  \label{eq:Coupl1}
\end{equation}%
The partial width of the decay $M\rightarrow D_{s}^{+}\pi ^{+}$ is
determined by the expression
\begin{equation}
\Gamma _{1}\left[ M\rightarrow D_{s}^{+}\pi ^{+}\right] =\frac{%
g_{1}^{2}m_{D_{s}}^{2}}{8\pi }\lambda \left( 1+\frac{\lambda ^{2}}{%
m_{D_{s}}^{2}}\right) ,  \label{eq:DW}
\end{equation}%
where $\lambda =\lambda (m,m_{D_{s}},m_{\pi })$ and
\begin{eqnarray}
\lambda \left( a,b,c\right) &=&\frac{\left[ a^{4}+b^{4}+c^{4}-2\left(
a^{2}b^{2}+a^{2}c^{2}+b^{2}c^{2}\right) \right] ^{1/2}}{2a}.  \notag \\
&&
\end{eqnarray}%
Then it is not difficult to find that
\begin{equation}
\Gamma _{1}\left[ M\rightarrow D_{s}^{+}\pi ^{+}\right] =(71\pm 23)~\mathrm{%
MeV},  \label{eq:DW1Numeric}
\end{equation}%
which is large enough to confirm dominant nature of this channel.

%%%%%%%%%%%%%%%%%%%%%%%%%%%%%%%%%%%%%%%%%%%%%%%%%%%%%%%%%%%%%%%%%%%%%%%%%%%%%

\subsection{Processes $M\rightarrow D_{s}^{\ast +}\protect\rho ^{+}$ and $%
M\rightarrow D^{\ast +}K^{\ast +}$}

%%%%%%%%%%%%%%%%%%%%%%%%%%%%%%%%%%%%%%%%%%%%%%%%%%%%%%%%%%%%%%%%%%%%%

These two processes differ from previous decay by a vector nature of
produced mesons. This fact modifies matrix elements of the vertices and
formulas for decay widths of the processes. We concentrate on analysis of
the decay $M\rightarrow D_{s}^{\ast +}\rho ^{+}$, and write down only the
final predictions for $M\rightarrow D^{\ast +}K^{\ast +}$.

The correlation function, which should be studied in the case of the decay $%
M\rightarrow D_{s}^{\ast +}\rho ^{+}$, has the following form%
\begin{equation}
\Pi _{\mu }(p,q)=i\int d^{4}xe^{ipx}\langle \rho (q)|\mathcal{T}\{J_{\mu
}^{D_{s}^{\ast }}(x)J^{\dag }(0)\}|0\rangle ,  \label{eq:CorrF7}
\end{equation}%
where $J_{\mu }^{D_{s}^{\ast }}(x)$ is the interpolating current for the
meson $D_{s}^{\ast +}$
\begin{equation}
J_{\mu }^{D_{s}^{\ast }}(x)=\overline{s}_{i}(x)\gamma _{\mu }c_{i}(x).
\label{eq:Dcur3}
\end{equation}%
The correlation function $\Pi _{\mu }(p,q)$ written down in terms of the
matrix elements of the particles and the vertex $MD_{s}^{\ast +}\rho ^{+}$
is given by the expression
\begin{eqnarray}
&&\Pi _{\mu }^{\mathrm{Phys}}(p,q)=\frac{\langle \rho (q,\epsilon
)D_{s}^{\ast +}(p,\varepsilon )|M(p^{\prime })\rangle \langle M(p^{\prime
})|J^{\dagger }|0\rangle }{(p^{\prime 2}-m^{2})}  \notag \\
&&\times \frac{\langle 0|J_{\mu }^{D_{s}^{\ast }}|D_{s}^{\ast
+}(p,\varepsilon )\rangle }{(p^{2}-m_{D_{s}^{\ast }}^{2})}+\cdots,
\label{eq:CorrF8}
\end{eqnarray}%
where $m_{D_{s}^{\ast }}$ and $\varepsilon _{\mu }$ are the mass and
polarization vector of the meson $D_{s}^{\ast +}$, and $\epsilon _{\nu }$ is
polarization vector of the $\rho $ meson. The expression in Eq.\ (\ref%
{eq:CorrF8}) is the contribution of the ground-state particles to the
physical side of the sum rule, whereas ellipses stand for the contributions
of higher resonances and continuum states.

We introduce the matrix element of the meson $D_{s}^{\ast +}$ by the formula%
\begin{equation}
\langle 0|J_{\mu }^{D_{s}^{\ast }}|D_{s}^{\ast +}(p,\varepsilon )\rangle
=m_{D_{s}^{\ast }}f_{D_{s}^{\ast }}\varepsilon _{\mu }.  \label{eq:Mel3}
\end{equation}%
We also model the mass-shell matrix element of the vertex $MD_{s}^{\ast
+}\rho ^{+}$ in the following way:%
\begin{eqnarray}
&&\langle \rho (q,\epsilon)D_{s}^{\ast +}(p,\varepsilon )|M(p^{\prime
})\rangle =g_{2} \left[ \left( q\cdot p\right) \left( \epsilon \cdot
\varepsilon ^{\ast }\right) \right.  \notag \\
&&\left. -\left( p\cdot \epsilon \right) \left( q\cdot \varepsilon ^{\ast
}\right) \right] ,  \label{eq:Ver}
\end{eqnarray}%
Then, it is not difficult to calculate the function $\Pi _{\mu }^{\mathrm{%
Phys}}(p,q)$,
\begin{eqnarray}
&&\Pi _{\mu }^{\mathrm{Phys}}(p,q)=\frac{fmm_{D_{s}^{\ast }}f_{D_{s}^{\ast }}%
}{(p^{\prime 2}-m^{2})(p^{2}-m_{D_{s}^{\ast }}^{2})}  \notag \\
&&\times \left( \frac{m_{\rho }^{2}+m_{D_{s}^{\ast }}^{2}-m^{2}}{2}\epsilon
_{\mu }+p\cdot \epsilon q_{\mu }\right) +\cdots ,  \label{eq:CorrF9}
\end{eqnarray}%
As is seen, $\Pi _{\mu }^{\mathrm{Phys}}(p,q)$ contains two structures
proportional to $\epsilon _{\mu }$ and $q_{\mu }$. We are going to employ
the structure $\sim \epsilon _{\mu }$, and corresponding invariant amplitude
which in the soft limit has the form
\begin{equation}
\widehat{\Pi }^{\mathrm{Phys}}(p^{2})=\frac{fmm_{D_{s}^{\ast
}}f_{D_{s}^{\ast }}}{(p^{2}-\widehat{m}^{2})^{2}}\frac{m_{\rho
}^{2}+m_{D_{s}^{\ast }}^{2}-m^{2}}{2},  \label{eq:CorrF9a}
\end{equation}%
with $\widehat{m}^{2}$ being equal to $(m^{2}+m_{D_{s}^{\ast }}^{2})/2$.

The QCD side of the sum rule is determined by the correlator%
\begin{eqnarray}
&&\Pi _{\mu }^{\mathrm{OPE}}(p,q)=i\int d^{4}xe^{ipx}\left[ \gamma _{\nu
}S_{s}^{bi}(-x){}\gamma _{\mu }S_{c}^{ia}(x)\gamma ^{\nu }\right] _{\alpha
\beta }  \notag \\
&&\times \langle \rho (q)|\overline{u}_{\alpha }^{b}(0)d_{\beta
}^{a}(0)|0\rangle.  \label{eq:QCDside3}
\end{eqnarray}%
This function has the same Lorentz structures as $\Pi _{\mu }^{\mathrm{Phys}%
}(p,q)$. In the soft-meson approximation, it receives contribution from the
matrix element
\begin{equation}
\langle 0|\overline{d}\gamma _{\nu }u|\rho \rangle =f_{\rho }m_{\rho
}\epsilon _{\nu },  \label{eq:Mel4}
\end{equation}%
where $f_{\rho }$ is the decay constant of the $\rho $ meson.

Having fixed an amplitude which is proportional to $\epsilon _{\mu }$ and
labeled it by $\widehat{\Pi }^{\mathrm{OPE}}(p^{2})$ it is not difficult to
write the sum rule for the strong coupling $g_{2}$
\begin{equation}
g_{2}=\frac{2\mathcal{P}(M^{2},\widehat{m}^{2})\widehat{\Pi }^{\mathrm{OPE}%
}(M^{2},s_{0})}{fmm_{D_{s}^{\ast }}f_{D_{s}^{\ast }}(m_{\rho
}^{2}+m_{D_{s}^{\ast }}^{2}-m^{2})}.  \label{eq:SCoupl2}
\end{equation}%
Here, $\widehat{\Pi }^{\mathrm{OPE}}(M^{2},s_{0})$ is the amplitude $%
\widehat{\Pi }^{\mathrm{OPE}}(p^{2})$ after the Borel transformation and
continuum subtraction procedures. It does not differ considerably from the $%
\Pi ^{\mathrm{OPE}}(M^{2},s_{0})$ and has the following form
\begin{eqnarray}
&&\widehat{\Pi }^{\mathrm{OPE}}(M^{2},s_{0})=\frac{f_{\rho }m_{\rho }}{48\pi
^{2}}\int_{\mathcal{M}^{2}}^{s_{0}}\frac{ds(m_{c}^{2}-s)}{s^{2}}  \notag \\
&&\times (m_{c}^{4}+m_{c}^{2}m_{s}-6m_{c}m_{s}s-2s^{2})e^{-s/M^{2}}  \notag
\\
&&+f_{\rho }m_{\rho }\widehat{\Pi }_{\mathrm{NP}}(M^{2}).  \label{eq:CorrF10}
\end{eqnarray}%
The nonperturbative term $\widehat{\Pi }_{\mathrm{NP}}(M^{2})$ is given by
the expression%
\begin{eqnarray}
&&\widehat{\Pi }_{\mathrm{NP}}(M^{2})=\frac{\langle \overline{s}s\rangle
m_{c}}{12M^{2}}\left( 2M^{2}-m_{c}m_{s}\right) e^{-m_{c}^{2}/M^{2}}  \notag
\\
&&+\langle \frac{\alpha _{s}G^{2}}{\pi }\rangle \frac{m_{c}}{144M^{4}}%
\int_{0}^{1}\frac{dx}{(1-x)^{3}}\left[ m_{c}^{3}x+(1-x)\left(
m_{c}^{2}m_{s}\right. \right.  \notag \\
&&\left. \left. -2m_{s}M^{2}-m_{c}M^{2}\right) \right]
e^{-m_{c}^{2}/[M^{2}(1-x)]}+\frac{\langle \overline{s}g\sigma Gs\rangle }{%
72M^{6}}  \notag \\
&&\times
(6m_{c}^{4}m_{3}-6m_{c}^{3}M^{2}-2m_{c}^{2}m_{s}M^{2}-m_{s}M^{4})e^{-m_{c}^{2}/M^{2}}
\notag \\
&&+\langle \frac{\alpha _{s}G^{2}}{\pi }\rangle \langle \overline{s}s\rangle
\frac{\pi ^{2}m_{c}}{36M^{8}}\left[
4m_{c}^{3}m_{s}-2M^{2}m_{c}(m_{c}+3m_{s})\right.  \notag \\
&&\left. +2M^{4}\right] e^{-m_{c}^{2}/M^{2}}+\langle \frac{\alpha _{s}G^{2}}{%
\pi }\rangle \langle \overline{s}g\sigma Gs\rangle \frac{\pi ^{2}m_{c}}{%
108M^{12}}  \notag \\
&&\times \left[ 9M^{6}-60m_{c}^{5}m_{s}+10m_{c}^{3}M^{2}(3m_{c}+7m_{s})%
\right.  \notag \\
&&\left. -4M^{4}m_{c}(9m_{c}+4m_{s})\right] e^{-m_{c}^{2}/M^{2}}.
\label{eq:CorrF11}
\end{eqnarray}

Numerical computations lead to the result
\begin{equation}
|g_{2}|=(9.8\pm 1.2)\times 10^{-1}~\mathrm{GeV}^{-1}.  \label{eq:Coupl2}
\end{equation}%
The partial width of the decay $M\rightarrow D_{s}^{\ast +}\rho ^{+}$ is
determined by the formula
\begin{equation}
\Gamma _{2}\left[ M\rightarrow D_{s}^{\ast +}\rho ^{+}\right] =\frac{%
|g_{2}|^{2}m_{\rho }^{2}}{8\pi }\widehat{\lambda }\left( 3+\frac{2\widehat{%
\lambda }^{2}}{m_{\rho }^{2}}\right),  \label{eq:DW2}
\end{equation}%
where $\widehat{\lambda }=\lambda (m,m_{D_{s}^{\ast }},m_{\rho })$. Then, we
get
\begin{equation}
\Gamma _{2}\left[ M\rightarrow D_{s}^{\ast +}\rho ^{+}\right] =(15\pm 4)~%
\mathrm{MeV}.  \label{eq:DW2Numeric}
\end{equation}

The decay $M\rightarrow D^{\ast +}K^{\ast +}$ can be considered in a similar
way. Omitting details, we write down predictions for the strong coupling $%
g_{3}$ and partial width of the process $M\rightarrow D^{\ast +}K^{\ast +}$:

\begin{equation}
g_{3}=(1.5\pm 0.3)~\mathrm{GeV}^{-1},  \label{eq:Coupl3}
\end{equation}%
and
\begin{equation}
\Gamma _{3}\left[ M\rightarrow D^{\ast +}K^{\ast +}\right] =(37\pm 10)~%
\mathrm{MeV}.  \label{eq:DW3Numeric}
\end{equation}

Information gained in this section allows us to compare couplings of the
current $J(x)$ to different two-meson states. It is seen that $g_{i}$
corresponding to vertices $MD^{\ast +}K^{\ast +}$, $MD_{s}^{\ast +}\rho ^{+}$
and $MD_{s}^{+}\pi ^{+}$ obey the inequalities $g_{3}>|g_{2}|>g_{1}$. Hence,
$J(x)$ describes mainly the hadronic molecule $D^{\ast +}K^{\ast +}$, as it
has been asserted in the Sec.\ \ref{sec:MassCoupl}. Differences in the
widths of relevant decays are connected not only with $g_{i}$, but generated
also by $\lambda $- factors and parameters of produced mesons.

Using results obtained for the partial widths of the decays considered in
this section, we can estimate the full width of the molecule $M$%
\begin{equation}
\Gamma =(123\pm 25)~\mathrm{MeV},  \label{eq:DWfull}
\end{equation}%
which should be confronted with the experimental data Eq.\ (\ref{eq:data1}).
It is seen that, although $\Gamma $ and $\Gamma _{\mathrm{exp}}$ do not
coincide, they are comparable with each other provided one takes into
account errors of the measurements and theoretical analyses.

\bigskip

\bigskip
%%%%%%%%%%%%%%%%%%%%%%%%%%%%%%%%%%%%%%%%%%%%%%%%%%%%%%%%%%%%%%%%%%%%%%%%%%%

\section{Conclusions}

\label{sec:Conclusions}
%%%%%%%%%%%%%%%%%%%%%%%%%%%%%%%%%%%%%%%%%%%%%%%%%%%%%%%%%%%%

In the present article, we have investigated features of the hadronic
molecule $M=D^{\ast +}K^{\ast +}$, and calculated its mass and width. The
mass of $M$ has been computed using QCD two-point sum rule method.
Prediction obtained for the mass $m=(2924~\pm 107)~\mathrm{MeV}$ is in 
nice agreement with the LHCb datum for the mass of the resonance $%
T_{cs0}^{a++}$. It also does not differ considerably  from the mass $%
(2917~\pm 135)~\mathrm{MeV}$ of the molecule $D_{s}^{\ast +}\rho ^{+}$
suggested to model $T_{cs0}^{a++}$ in our paper \cite{Agaev:2022duz} .

We have evaluated the width of the molecule $M$ by calculating partial
widths of the decay channels $M\rightarrow D_{s}^{+}\pi ^{+}$, $M\rightarrow
D_{s}^{\ast +}\rho ^{+}$, and $M\rightarrow D^{\ast +}K^{\ast +}$. To this
end, we have found strong couplings of the particles at vertices $%
MD_{s}^{+}\pi ^{+}$, $MD_{s}^{\ast +}\rho ^{+}$, and $MD^{\ast +}K^{\ast +}$
by means of QCD light-cone sum rule approach and soft-meson approximation.
The strong coupling of the hadronic molecule $M$ with final-state mesons is
large in the case of mesons $D^{\ast +}$ and $K^{\ast +}$, which is
understandable because $M$ is composed of these particles. Nevertheless, the
partial width of the decay $M\rightarrow D^{\ast +}K^{\ast +}$ is less than
that of the decay $M\rightarrow D_{s}^{+}\pi ^{+}$. The reason is that the
kinematical factor $\lambda $ in the expression of the decay widths in Eqs.\
(\ref{eq:DW}) and (\ref{eq:DW2}) gets its largest value in the process $%
M\rightarrow D_{s}^{+}\pi ^{+}$. As a result, the dominant channel of $M$ is
the decay to a pair of the mesons $D_{s}^{+}$ and $\pi ^{+}$, which was
actually observed by the LHCb collaboration.

The final result $\Gamma =(123\pm 25)~\mathrm{MeV}$ for the full width of
the molecule $M$ is compatible with the $\Gamma _{\mathrm{exp}}$ within the
existing experimental and theoretical errors. The estimate for $\Gamma $ may
be further improved by taking into account another decay channels of the
hadronic molecule $M$. It will be interesting also to compare the $\Gamma $
and $\Gamma _{\mathrm{exp}}$ with predictions for the full width of the
resonance $T_{cs0}^{a++}$ obtained in the context of alternative methods.
Nevertheless, based on our present results, we may consider the hadronic
molecule $M=D^{\ast +}K^{\ast +}$ as a possible candidate to the doubly
charged resonance $T_{cs0}^{a++}$.

The isoscalar partner of $T_{cs0}^{a++}$, namely the second resonance $%
T_{cs0}^{a0}$ with quark content $cd\overline{s}\overline{u}$ may be modeled
as a linear superposition of hadronic molecules $D^{\ast 0}K^{\ast 0}$ and $%
D_{s}^{\ast +}\rho ^{-}$. The dominant decay channel of this state is the
process $T_{cs0}^{a0}\rightarrow D_{s}^{+}\pi ^{-}$. The modes $%
T_{cs0}^{a0}\rightarrow D^{0}K^{0}$, $D_{s}^{\ast +}\rho ^{-}$ and $D^{\ast
0}K^{\ast 0}$ are other possible decay channels of $T_{cs0}^{a0}$.
Experimentally measured mass and width differences between $T_{cs0}^{a++}$
and $T_{cs0}^{a0}$ are equal to $\Delta m\approx 28~\mathrm{MeV}$ and $%
\Delta \Gamma \approx 15~\mathrm{MeV}$, respectively. To be accepted as a
reliable model for $T_{cs0}^{a0}$ the molecule picture should be
successfully confronted with the available data.

Another problem to be addressed here, is similarities and differences of the
resonances $T_{cs0}^{a0/++}$ and $X_{0(1)}$. It has been noted in Sec.\ \ref%
{sec:Intro} that $X_{0(1)}$ are exotic mesons composed of quarks $ud%
\overline{s}\overline{c}$. The scalar structures $T_{cs0}^{a0}$ and $X_{0}$
differ from each other by quark-exchanges $\overline{u}\leftrightarrow u$
and $c\leftrightarrow \overline{c}$. They are neutral particles with masses $%
2892~\mathrm{MeV}$ and $2866~\mathrm{MeV}$, respectively. It is seen, that a
mass gap between these two structures is small. In Ref.\ \cite{Agaev:2020nrc}%
, we modeled $X_{0}$ as a hadronic molecule $\overline{D}^{\ast 0}K^{\ast 0}$
and found its mass equal to $2868~\mathrm{MeV}$, which is in a very nice
agreement with the LHCb datum. The $\overline{D}^{\ast 0}K^{\ast 0}$ and a
component $D^{\ast 0}K^{\ast 0}$ of the molecule model for $T_{cs0}^{a0}$,
have almost identical structures, therefore one expects the mass of the
isoscalar state $T_{cs0}^{a0}$ will be be consistent with experiments.

The mass difference $\sim 55~\mathrm{MeV}$ between $T_{cs0}^{a++}$ and $%
X_{0} $ is larger than the gap in the previous case which is connected
presumably with doubly charged nature of $T_{cs0}^{a++}$. We explored the
vector resonance $X_{1}$ in Ref.\ \cite{Agaev:2021knl} as a
diquark-antidiquark state $[ud][\overline{c}\overline{s}]$ and achieved
reasonable agreements with the LHCb data. In general, hadronic molecules
composed of two mesons may be used to model vector particles as well.
Because $T_{cs0}^{a++}$ has the spin-parity $J^{\mathrm{P}}=0^{+}$, in this
article, we have studied only scalar particle $D^{\ast +}K^{\ast +}$.
Molecules with the same quark content but different spin-parities are yet
hypothetical structures. They are interesting objects for theoretical
researches as well, because may be discovered soon in various exclusive
processes. Properties of the isoscalar resonance $T_{cs0}^{a0}$, as well as
counterparts of $T_{cs0}^{a0/++}$ with different spin-parities are issues
for future investigations.

\end{document}